\documentclass[aps, prb, preprint, superscriptaddress,longbibliography]{revtex4-1}
\usepackage{graphicx}
\usepackage{bm}
\usepackage{natbib}
\usepackage{upgreek}
\usepackage{amsmath}
\usepackage{amssymb}

\begin{document}

\title{Superradiant and transport lifetimes of the cyclotron resonance in the topological insulator HgTe}

\author{J. Gospodari\v{c}}
\author{V. Dziom}
\author{A. Shuvaev}
\affiliation{Institute of Solid State Physics, Vienna University of
Technology, 1040 Vienna, Austria}
\author{A. A. Dobretsova}
\author{N. N. Mikhailov}
\author{Z. D. Kvon}
\affiliation{Rzhanov Institute of Semiconductor Physics and Novosibirsk State University, Novosibirsk 630090, Russia}
\author{A. Pimenov}
\affiliation{Institute of Solid State Physics, Vienna University of
Technology, 1040 Vienna, Austria}

\begin{abstract}

We investigate the superradiance effects in three-dimensional topological insulator HgTe with conducting surface states.  We demonstrate that the superradiance can be explained using the classical electrodynamic approach. Experiments using the continuous-wave spectroscopy allowed to separate the energy losses in the system into intrinsic and radiation losses, respectively. These results demonstrate that the superradiance effects are not sensitive to the details of the band structure of the material.

\end{abstract}

\date{\today}


\maketitle

\subsection{Introduction}

The phenomenon of superradiance occurs when several emitters become coupled and, as a consequence, they start to radiate coherently~\cite{dicke_pr_1954, rehler_pra_1971}. In such a case, the intensity of the emitted radiation is strongly enhanced as it is scaled as a square of the number of sources in contrast to simple proportionality without coherence. Superradiance was first predicted theoretically by Dicke~\cite{dicke_pr_1954} for an ensemble of two-level atoms and has been later on observed experimentally in various materials\cite{skribanowitz_prl_1973, gross_prl_1979, kaluzny_prl_1983} and metamaterials\cite{sonnefraud_nano_2010, wenclawiak_apl_2017},  see Ref.~[\onlinecite{cong_josab_2016}] for a review.

Recently, superradiant effects in two-dimensional electron gases (2DEGs) have been brought into attention~\cite{zhang_prl_2014, laurent_prl_2015, zhang_np_2016, curtis_prb_2016, maag_np_2016, muravev_prb_2017, moiseev_prb_2017}. In the original Dicke model, the coherence is obtained because single sources are within the distance smaller than the wavelength of the radiation. An important difference especially compared to the case of the cyclotron resonance in 2DEGs is that the size limitation is lifted in the latter case. The resonance of a single electron is excited coherently by an incident electromagnetic wave. Therefore, independently of the sample size, they re-emit a coherent secondary wave. We recall that the consideration of the secondary waves is a standard procedure to calculate the interaction between electromagnetic waves with matter~\cite{jackson_book}. In addition to the radiative losses, usual scattering processes, like impurity scattering, contribute to the lifetime of the cyclotron resonance. In 2DEGs the effects of the superradiance may be easily monitored experimentally, e.g. by varying the temperature or gate voltage.

The main experimental procedure to investigate the superradiance in 2DEGs is based on the time-domain terahertz spectroscopy~\cite{zhang_prl_2014, zhang_np_2016, curtis_prb_2016, maag_np_2016, moiseev_prb_2017}. Within this technique~\cite{nuss_book} the cyclotron resonance is excited by a short terahertz pulse and the free decay of the electronic system is directly recorded in the time domain. Here we utilize an alternative approach of a continuous-wave spectroscopy~\cite{kozlov_book} to the phenomenon of superradiance. This method is applied to the three-dimensional topological insulator HgTe, where the electronic band structure deviates from classical parabola due to the strong spin-orbit coupling.

Three-dimensional topological insulators (TI) represent a class of materials that are insulating in the bulk but reveal conducting two-dimensional surface states~\cite{hasan_rmp_2010,qi_rmp_2011}. The surface states in TI have a non-degenerate Dirac-like dispersion with electron spin locked to the direction of the momentum. In the 3D TIs based on HgTe quantum wells the inversion of the $\Gamma_6$ and $\Gamma_8$ bands in the dispersion of the bulk HgTe leads to Dirac-like surface states at the interface. Applying strain with the CdTe/HgTe/CdTe structure forms an insulating gap~\cite{brune_prl_2011,Kozlov2014} (above $10$~meV) between the light-hole and heavy-hole $\Gamma_8$ bands, making the strained HgTe a 3D TI. Since the Fermi level of the ungated sample lies in the bulk band gap between the light-hole conduction and heavy-hole valence band, the electrodynamics of the system is governed only by a 2D surface states, with negligible effects from the bulk carriers at low-temperatures~\cite{brune_prl_2011,shuvaev_prb_2013}. The Dirac-point of the surface states is presumed to be located deep below the heavy-hole band~\cite{dziom_ncomm_2017}.

\subsection{Experimental}\label{secexp}

The experiments were carried out on a strained $80$\,nm thick mercury telluride ({HgTe}) layer sandwiched between the thin cap (top) and buffer (bottom) layers of Cd$_{0.3}$Hg$_{0.7}$Te. The introduction of the indicated buffer layer in such samples results in a high electron mobility of $4\times10^5$\,cm$^{2}$/Vs, which is the current record value for a 3D TI \cite{Kozlov2014}. The quantum-well structure was grown by molecular beam epitaxy on a (100)-oriented {GaAs} substrate\cite{kvon_ltp_2009}. The film was covered on top by a multilayer insulator of ${SiO_{2}}/{Si_3N_4}$ and a metallic 10.5\,nm thick Ti-Au layer, which acts as a semitransparent top-gate electrode and allowed the variation of the charge density~\cite{shuvaev_apl_2013, kozlov_prl_2016}.

The sample was investigated in a Mach-Zehnder interferometer~\cite{volkov_infrared_1985}, which allowed us to acquire the amplitude and the phase shift of the transmitted electromagnetic radiation in parallel and crossed polarizer geometries~\cite{shuvaev_sst_2012, shuvaev_apl_2013}.  Continuous monochromatic radiation was produced by a backward-wave oscillators operating in the submillimeter regime ($10^2$\,GHz\,--\,$10^3$\,GHz). The experimental results in this work were obtained at a fixed frequency of 208\,GHz in sweeping magnetic fields. Additional information about the charge carriers was also obtained from the frequency-dependent spectra in zero magnetic field. Transmission experiments were carried out at $2$ K in a split-coil superconducting magnet that provided an external magnetic field up to $7$\,T in the Faraday geometry; i.e., a magnetic field was applied along the propagation direction of the radiation.

\subsection{Spectra modelling}\label{thrmod}

As extensively discussed previously, the superradiance effects in 2DEGs may be well explained via the classical picture~\cite{matov_jetp_1996, mikhailov_prb_1996, mikhailov_prb_2004}. Indeed, although the cyclotron resonance is the transition between quantized Landau levels, for not too high fields a quasi-classical approach is sufficient~\cite{ashcroft_book} because several levels take part in the transition process~\cite{shuvaev_prb_2017}. In this case the cyclotron resonance is scaled linearly with the magnetic field regardless of the details of the band structure. In the following, we reproduce briefly the main expressions describing the cyclotron resonance in 2DEGs including the effects of the radiation losses (superradiance).

We consider a geometry with the external magnetic field and the propagation direction of the incident wave being perpendicular to the film surface. The film is assumed to be thin compared to the radiation wavelength inside the sample and, finally, in this section we neglect the effect of the substrate for simplicity. We note that explicitly including the dielectric substrate into account\cite{shuvaev_sst_2012, dziom_2d_2017, dziom_ncomm_2017} (see Section~\ref{sechgte}) produces only marginal changes in the spectra.

The classical equation of motion for an electron in a static magnetic field can be written as:
\begin{equation}\label{eqmotion}
  d\mathbf{v}/dt +\frac{1}{\tau_0}\mathbf{v} - \frac{e}{m} \mathbf{v}\times \mathbf{B} = \frac{e}{m} \mathbf{E}_{int}e^{-i\omega t} . \
\end{equation}
Here, $\mathbf{B}$ is the static external magnetic field, $\mathbf{E}_{int}$ is the $ac$ electric field inside the sample, $e$ and $m$ are the electron charge and cyclotron mass, respectively, and $\tau_0$ is the usual transport scattering time that does not include the radiation losses. As usual for continuous wave approach, we assume $e^{-i\omega t}$ time dependence of all $ac$ fields. Note that Eq.~(\ref{eqmotion}) leads to the standard Drude conductivity in magnetic fields; the diagonal and the Hall components of the conductivity tensor can be written as:
\begin{equation}\label{drudecondxx}
  \sigma_{xx}(\omega)=\sigma_{yy}(\omega)=\frac{1-i\omega\tau_0}{(1-i\omega\tau_0)^2+(\Omega_c\tau_0)^2}\sigma_0
\end{equation}
\begin{equation}\label{drudecondxy}
  \sigma_{xy}(\omega)=-\sigma_{yx}(\omega)=\frac{\Omega_c\tau_0}{(1-i\omega\tau_0)^2+(\Omega_c\tau_0)^2}\sigma_0 \ .
\end{equation}
Here, $\Omega_c = eB/m$ gives the cyclotron resonance frequency, $\sigma_0=ne^2\tau_0/m$ is the 2D static conductivity and $n$ is the 2D carrier density. The conductivity tensor $\sigma$ connects the 2D current in the film, $\mathbf{j}_{2D}$, with $\mathbf{E}_{int}$ as: $\mathbf{j}_{2D}=\sigma\mathbf{E}_{int}$.

\begin{figure}[tbp]
    \centering{}\includegraphics[width=0.7\linewidth, clip]{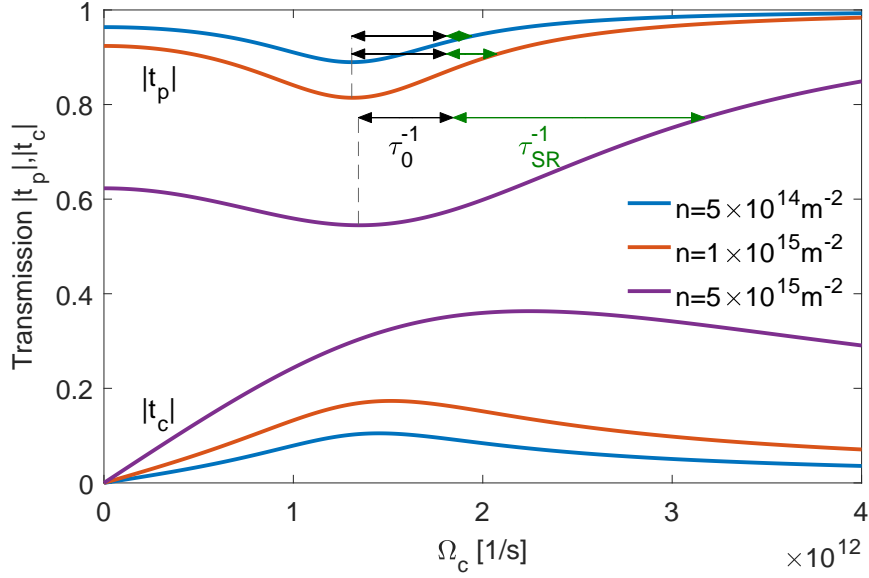}\caption{Calculated amplitudes of parallel $|t_p|$ and crossed $|t_c|$ magneto-optical transmission at $\nu=208$\,GHz using Eqs.~(\ref{eqtpp},\ref{eqtpc}). The spectra corresponds to a thin film with electronic carriers with the cyclotron mass $m=0.02\,m_0$, scattering time $\tau_0=2\times10^{-12}$\,s, and with varied density as indicated. The apparent width of the cyclotron resonance is determined by $\Gamma = {1}/{\tau_0}+ {1}/{\tau_{\mathrm{SR}}}$ (see text). At low 2D densities $n$ the intrinsic scattering ${1}/{\tau_0}$ determines the resonance width. By increasing the density the superradiant decay becomes the dominating mechanism for the energy loss in the film.}
     \label{fig:figtheory}
\end{figure}

In order to explicitly include the radiation losses, $\mathbf{E}_{int}$ must be connected to the fields of the electromagnetic waves outside the film.  In the thin film approximation the electric field may be considered as homogeneous across the sample, i.e. $\mathbf{E}_0 + \mathbf{E}_r = \mathbf{E}_{int} = \mathbf{E}_t$. Here, $\mathbf{E}_0$ is the field of the incident wave, which is taken as linearly polarized, and $\mathbf{E}_r, \mathbf{E}_t$ are the reflected and transmitted waves, respectively. The boundary condition for the magnetic fields in the case of thin conducting film~\cite{dressel_book_2002} is obtained via Maxwell equation $\oint \mathbf{H} d\mathbf{s} =  \iint (\partial \mathbf{D}/\partial t) d\mathbf{A}$ leading to the modified boundary condition for tangential magnetic fields on the left and right sides of the film, respectively: $\mathbf{H}_l - \mathbf{H}_r = \sigma \mathbf{E}_{int} = \mathbf{j}_{2D} = ne\mathbf{v}$. Substituting both boundary conditions into Eq.\,(\ref{eqmotion}) we obtain a modified equation of motion as a function of the incident wave:
\begin{equation}\label{eqmotion2}
  d\mathbf{v}/dt +(\frac{1}{\tau_0}+\frac{1}{\tau_{\mathrm{SR}}})\mathbf{v} - \frac{e}{m} \mathbf{v}\times \mathbf{B} = \frac{e}{m} \mathbf{E}_{0}e^{-i\omega t}  \ ,
\end{equation}
where $1/\tau_{\mathrm{SR}} = ne^2 Z_0/2m$ now takes into account the radiation losses explicitly. Here, $Z_0$ is the impedance of the free space. Equation~(\ref{eqmotion2}) demonstrates that the losses in a thin film can be represented as a sum of two contributions, given by the transport and radiative lifetimes.

After a usual algebra, Eq.~(\ref{eqmotion2}) leads to well-known expressions for transmission through the film~\cite{shuvaev_sst_2012, dziom_2d_2017, dziom_ncomm_2017} that, e.g., in the geometry with parallel and crossed polarizers can be written as:
\begin{equation}\label{eqtpp}
  t_p = 1 - \frac{i}{\tau_{\mathrm{SR}}}\frac{\omega+i \Gamma}{(\omega+i \Gamma)^2-\Omega_c^2} \ ,
\end{equation}
\begin{equation}\label{eqtpc}
  t_c = \frac{1}{\tau_{\mathrm{SR}}}\frac{\Omega_c}{\Omega_c^2-(\omega+i \Gamma)^2} \ ,
\end{equation}
where $\Gamma = {1}/{\tau_0}+ {1}/{\tau_{\mathrm{SR}}}$ is the "total" scattering rate.

Figure~\ref{fig:figtheory} shows the calculated parallel and crossed transmission spectra of a conductive film with typical parameters ($\tau_0=2\times10^{-12}$\,s, $m=0.02\,m_0$) at the frequency $\nu=208$\,GHz and for varied electron densities. This frequency was taken to match that of the experiments. It is clear that the nominators in Eq.~(\ref{eqtpp}) and Eq.~(\ref{eqtpc})  lead to a resonance-like form of the transmission spectra at $\omega = \Omega_c$ and with the width determined by $\tau_0$ and $\tau_{\mathrm{SR}}$. Here, the electron density $n$ was the only varied parameter, while others, including the scattering time $\tau_0$, were fixed. With the help of green and black arrows in  Fig.~\ref{fig:figtheory}, indicating the amplitude of inverse $\tau_{\mathrm{SR}}$ and $\tau_0$, we can see a direct correlation between the width of the resonance with the "total" scattering rate $\Gamma$. At low electron densities $n$ the total scattering rate and, consequently, the resonance width is characterized by intrinsic losses only. With increasing $n$,  the energy loss of the system becomes more and more dominated by radiative losses (see data corresponding to $n=5\times10^{15}\text{ m}^{-2}$ in Fig.~\ref{fig:figtheory}). It is important to note that the radiative losses $1/\tau_{\mathrm{SR}} = ne^2 Z_0/2m$ do not add further free parameters to the experiment, because the electron density and the cyclotron mass are independently determined via fitting the transmission spectra as described in the next section.

\subsection{Superradiance in HgTe}\label{sechgte}

\begin{figure}[tbp]
    \centering{}\includegraphics[width=0.7\linewidth, clip]{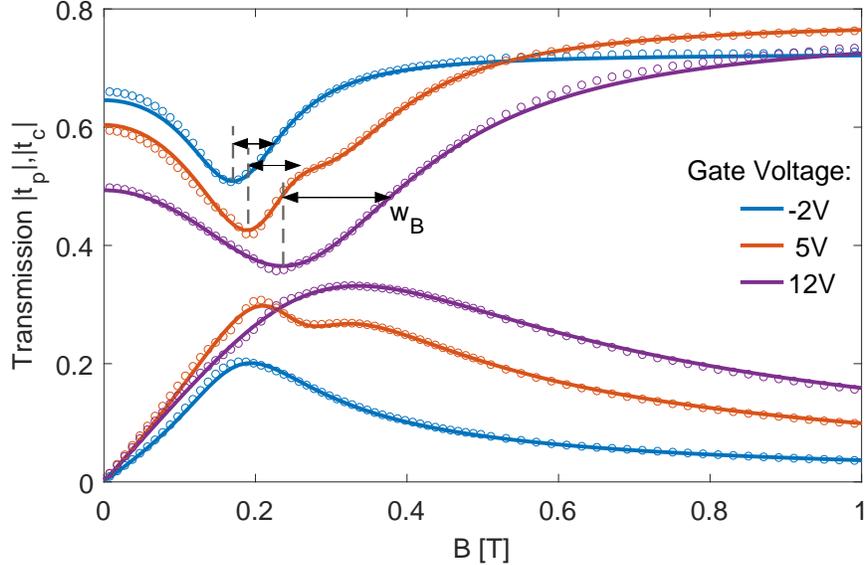}\caption{Experimental amplitudes of parallel $|t_p|$ and crossed $|t_c|$ magneto-optical transmission at  $\nu=208$\,GHz through {HgTe} film on a {GaAs} substrate and for different gate voltages. Symbols: experimental data, lines: model fit using the Drude model. Black arrows and $w_\mathrm{B}$ represent the apparent width of the resonance curves.}
     \label{fig:figmeas}
\end{figure}

The implementation of the transparent metallic gate on top of the {HgTe} film enabled us to systematically vary the 2D charge density in this system~\cite{shuvaev_apl_2013, shuvaev_prl_2016, shuvaev_prb_2017}. The voltage applied to the gate ranged between $-2$\,V and $+12$\,V and resulted in magneto-optical transmission spectra shown in Fig.~\ref{fig:figmeas}.

The magnetic field-dependence in the whole gate voltage range is dominated by a strong cyclotron resonance of the Dirac-like surface electrons located in the band gap\cite{shuvaev_sst_2012}. These states are mainly responsible for the Faraday rotation of the incident radiation. With increasing the gate voltage the response of such states becomes stronger, and the resonance is getting broader. These spectra qualitatively resemble that in Fig.~\ref{fig:figtheory}, where the broadening is due to increased density only.  In addition, for the gate voltage range between $1.5$\,V and $9.0$\,V we observed the appearance of a second weaker signal (see double resonance response at $5$\,V in Fig.~\ref{fig:figmeas}) that corresponds to effective masses of about $0.033\text{m}_{0}$. According to the band structure calculation for the 3D HgTe films\,\cite{dziom_ncomm_2017}, we attribute this effect to the bulk conduction band. Indeed, this effective mass agrees well with the cyclotron mass of thick unstrained HgTe films\,\cite{Berchenko1976,shuvaev_sst_2012} ($m \approx 0.03\,m_e$).
As the additional signal is weak, its properties do not affect the present discussion.

\begin{figure}[tbp]
    \centering{}\includegraphics[width=0.7\linewidth, clip]{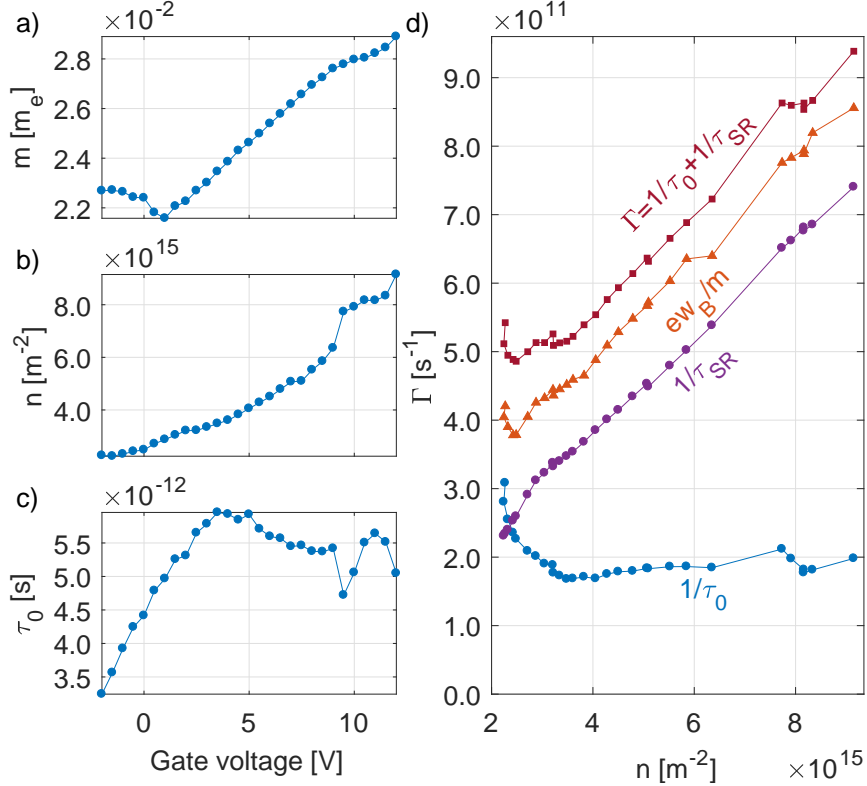}\caption{Parameters of the surface states obtained from fitting the experimental data in Fig.~\ref{fig:figmeas}. The effective mass $m^{\ast}$ (a), 2D density $n$ (b) and the transport scattering time constant $\tau_0$ (c) are plotted with respect to the applied gate voltage. (d) Solid circles: transport scattering rate $1/\tau_0$ and radiative scattering rate obtained via $1/\tau_{\mathrm{SR}}= ne^2 Z_0/(1+n_{\mathrm{CdTe}})m$. Solid squares: "total" scattering rate $\Gamma$ as compared to the scattering rate calculated directly from the estimated widths $w_\mathrm{B}$ of the resonances in Fig.~\ref{fig:figmeas} (solid triangles).}
     \label{fig:figpar}
\end{figure}

The measured transmission spectra can be fitted well within a Drude model, Eqs.~(\ref{drudecondxx},\ref{drudecondxy}). From the conductivity tensor one can obtain an explicit analytical formula for the transmission matrix for the case of a thin film on a substrate that accounts for the Fabry-P\'{e}rot interferences in the substrate (see Refs.\,[\onlinecite{dziom_2d_2017, dziom_ncomm_2017, shuvaev_prb_2017}] for exact expressions). For each applied gate voltage we simultaneously fit the field dependencies of the amplitude and the phase shift of both, parallel and crossed transmissions, where $n$, $\tau_0$ and $m$, characterized by the surface states, were set as free fitting parameters.

The electrodynamic parameters of the surface charge carriers obtained from the fitting of the magneto-optical data are shown in Fig.~\ref{fig:figpar}. As may be expected, the electron density of the surface states in Fig.~\ref{fig:figpar}(b) roughly linearly increases with the gate voltage. On the contrary, the cyclotron mass and the intrinsic scattering time $\tau_0$ show much weaker changes. The surface cyclotron mass of about $0.027$\,$m_e$ is smaller but similar to the bulk mass in HgTe\,\cite{shuvaev_sst_2012}. This similarity is due to strong hybridisation of the bulk and surface states\,\cite{brune_prl_2011}.

The intrinsic scattering time $\tau_0$ experiences a maximum as a function of the applied gate voltage, which we attribute to the competition of two scattering mechanisms \cite{Tkachov2011, Savchenko2018}. At first, increasing applied gate voltage causes scattering to decrease (increase of $\tau_0$) due to the screening of impurities by higher electron density. The decrease of $\tau_0$ at higher densities (gate voltage above $5$~V), where the impurity scattering weakens, is associated with the increasing scattering on the fluctuations of the quantum well width. The resulting fluctuations of the band gap create mass disorder for Dirac-like states in the system. The material parameters in Fig.~\ref{fig:figpar}\,(a-c) are sufficient to calculate the radiative losses via $1/\tau_{\mathrm{SR}} = ne^2 Z_0/(1+n_{\mathrm{CdTe}})m$, a modified version of the formula presented in Section~\ref{thrmod}, which also considers the HgTe film being enclosed by vacuum on one side and the CdTe substrate with the refractive index of $n_{\mathrm{CdTe}}=3.54$ on the other side. The values thus obtained are plotted in Fig.~\ref{fig:figpar}\,(d) as a function of the electron density. We see that in the present experiment two regimes can be obtained using the gate voltage, one being below $n\approx3\times10^{15}\text{ m}^{-2}$ with a comparable contribution from both mechanisms to the system losses ($\tau_0^{-1} \sim \tau_{\mathrm{SR}}^{-1}$) and the other at higher electron densities, where the radiative losses are dominant $\tau_0^{-1} < \tau_{\mathrm{SR}}^{-1}$.

We recall that the width of the resonance curves (Figs.\,\ref{fig:figtheory},\,\ref{fig:figmeas}) depends upon the total scattering rate $\Gamma$. This fact is confirmed in Fig.~\ref{fig:figpar}\,(d) directly comparing the total scattering rate $\Gamma$ (squares) with the width of the resonance $e w_\mathrm{B}/m$ (triangles). We note a good agreement between both data sets. These data demonstrate that internal and radiative losses can be seen in the continuous-wave spectra directly. Finally, in the present experiments, the obvious broadening of the observed cyclotron resonance curves is purely due to the increase of the radiative losses.

\subsection{Conclusions}

Using terahertz magneto-spectroscopy we investigated the cyclotron resonance in a three-dimensional topological insulator HgTe with conducting surface states. From the analysis of the complex transmission coefficients the radiative and transport lifetimes can be well separated in the continuous-wave spectra. We have shown experimentally that at high carrier densities the super-resonant radiation dominates the energy losses in the system and they can be well explained via a classical electrodynamic picture. Within this approach, the coherent emission is established via the coherent interaction of the incident radiation with a thin-film sample that is not sensitive to the details of the band structure. These results demonstrate that the superradiance in semiconducting thin films should be interpreted as a fully classical effect.

\subsection*{Acknowledgments}

This work was supported by Austrian Science Funds (W-1243, P27098-N27, I3456-N27) and by Russian Foundation for Basic Research (17-52-14007).


\bibliography{literature}

\begin{thebibliography}{40}%
	\makeatletter
	\providecommand \@ifxundefined [1]{%
		\@ifx{#1\undefined}
	}%
	\providecommand \@ifnum [1]{%
		\ifnum #1\expandafter \@firstoftwo
		\else \expandafter \@secondoftwo
		\fi
	}%
	\providecommand \@ifx [1]{%
		\ifx #1\expandafter \@firstoftwo
		\else \expandafter \@secondoftwo
		\fi
	}%
	\providecommand \natexlab [1]{#1}%
	\providecommand \enquote  [1]{``#1''}%
	\providecommand \bibnamefont  [1]{#1}%
	\providecommand \bibfnamefont [1]{#1}%
	\providecommand \citenamefont [1]{#1}%
	\providecommand \href@noop [0]{\@secondoftwo}%
	\providecommand \href [0]{\begingroup \@sanitize@url \@href}%
	\providecommand \@href[1]{\@@startlink{#1}\@@href}%
	\providecommand \@@href[1]{\endgroup#1\@@endlink}%
	\providecommand \@sanitize@url [0]{\catcode `\\12\catcode `\$12\catcode
		`\&12\catcode `\#12\catcode `\^12\catcode `\_12\catcode `\%12\relax}%
	\providecommand \@@startlink[1]{}%
	\providecommand \@@endlink[0]{}%
	\providecommand \url  [0]{\begingroup\@sanitize@url \@url }%
	\providecommand \@url [1]{\endgroup\@href {#1}{\urlprefix }}%
	\providecommand \urlprefix  [0]{URL }%
	\providecommand \Eprint [0]{\href }%
	\providecommand \doibase [0]{http://dx.doi.org/}%
	\providecommand \selectlanguage [0]{\@gobble}%
	\providecommand \bibinfo  [0]{\@secondoftwo}%
	\providecommand \bibfield  [0]{\@secondoftwo}%
	\providecommand \translation [1]{[#1]}%
	\providecommand \BibitemOpen [0]{}%
	\providecommand \bibitemStop [0]{}%
	\providecommand \bibitemNoStop [0]{.\EOS\space}%
	\providecommand \EOS [0]{\spacefactor3000\relax}%
	\providecommand \BibitemShut  [1]{\csname bibitem#1\endcsname}%
	\let\auto@bib@innerbib\@empty
	\bibitem [{\citenamefont {Dicke}(1954)}]{dicke_pr_1954}%
	\BibitemOpen
	\bibfield  {author} {\bibinfo {author} {\bibfnamefont {R.~H.}\ \bibnamefont
			{Dicke}},\ }\bibfield  {title} {\enquote {\bibinfo {title} {Coherence in
				spontaneous radiation processes},}\ }\href {\doibase 10.1103/PhysRev.93.99}
	{\bibfield  {journal} {\bibinfo  {journal} {Phys. Rev.}\ }\textbf {\bibinfo
			{volume} {93}},\ \bibinfo {pages} {99--110} (\bibinfo {year}
		{1954})}\BibitemShut {NoStop}%
	\bibitem [{\citenamefont {Rehler}\ and\ \citenamefont
		{Eberly}(1971)}]{rehler_pra_1971}%
	\BibitemOpen
	\bibfield  {author} {\bibinfo {author} {\bibfnamefont {N.~E.}\ \bibnamefont
			{Rehler}}\ and\ \bibinfo {author} {\bibfnamefont {J.~H.}\ \bibnamefont
			{Eberly}},\ }\bibfield  {title} {\enquote {\bibinfo {title}
			{Superradiance},}\ }\href {\doibase 10.1103/PhysRevA.3.1735} {\bibfield
		{journal} {\bibinfo  {journal} {Phys. Rev. A}\ }\textbf {\bibinfo {volume}
			{3}},\ \bibinfo {pages} {1735--1751} (\bibinfo {year} {1971})}\BibitemShut
	{NoStop}%
	\bibitem [{\citenamefont {Skribanowitz}\ \emph {et~al.}(1973)\citenamefont
		{Skribanowitz}, \citenamefont {Herman}, \citenamefont {MacGillivray},\ and\
		\citenamefont {Feld}}]{skribanowitz_prl_1973}%
	\BibitemOpen
	\bibfield  {author} {\bibinfo {author} {\bibfnamefont {N.}~\bibnamefont
			{Skribanowitz}}, \bibinfo {author} {\bibfnamefont {I.~P.}\ \bibnamefont
			{Herman}}, \bibinfo {author} {\bibfnamefont {J.~C.}\ \bibnamefont
			{MacGillivray}}, \ and\ \bibinfo {author} {\bibfnamefont {M.~S.}\
			\bibnamefont {Feld}},\ }\bibfield  {title} {\enquote {\bibinfo {title}
			{Observation of {D}icke superradiance in optically pumped {HF} gas},}\ }\href
	{\doibase 10.1103/PhysRevLett.30.309} {\bibfield  {journal} {\bibinfo
			{journal} {Phys. Rev. Lett.}\ }\textbf {\bibinfo {volume} {30}},\ \bibinfo
		{pages} {309--312} (\bibinfo {year} {1973})}\BibitemShut {NoStop}%
	\bibitem [{\citenamefont {Gross}\ \emph {et~al.}(1979)\citenamefont {Gross},
		\citenamefont {Goy}, \citenamefont {Fabre}, \citenamefont {Haroche},\ and\
		\citenamefont {Raimond}}]{gross_prl_1979}%
	\BibitemOpen
	\bibfield  {author} {\bibinfo {author} {\bibfnamefont {M.}~\bibnamefont
			{Gross}}, \bibinfo {author} {\bibfnamefont {P.}~\bibnamefont {Goy}}, \bibinfo
		{author} {\bibfnamefont {C.}~\bibnamefont {Fabre}}, \bibinfo {author}
		{\bibfnamefont {S.}~\bibnamefont {Haroche}}, \ and\ \bibinfo {author}
		{\bibfnamefont {J.~M.}\ \bibnamefont {Raimond}},\ }\bibfield  {title}
	{\enquote {\bibinfo {title} {Maser oscillation and microwave superradiance in
				small systems of rydberg atoms},}\ }\href {\doibase
		10.1103/PhysRevLett.43.343} {\bibfield  {journal} {\bibinfo  {journal} {Phys.
				Rev. Lett.}\ }\textbf {\bibinfo {volume} {43}},\ \bibinfo {pages} {343--346}
		(\bibinfo {year} {1979})}\BibitemShut {NoStop}%
	\bibitem [{\citenamefont {Kaluzny}\ \emph {et~al.}(1983)\citenamefont
		{Kaluzny}, \citenamefont {Goy}, \citenamefont {Gross}, \citenamefont
		{Raimond},\ and\ \citenamefont {Haroche}}]{kaluzny_prl_1983}%
	\BibitemOpen
	\bibfield  {author} {\bibinfo {author} {\bibfnamefont {Y.}~\bibnamefont
			{Kaluzny}}, \bibinfo {author} {\bibfnamefont {P.}~\bibnamefont {Goy}},
		\bibinfo {author} {\bibfnamefont {M.}~\bibnamefont {Gross}}, \bibinfo
		{author} {\bibfnamefont {J.~M.}\ \bibnamefont {Raimond}}, \ and\ \bibinfo
		{author} {\bibfnamefont {S.}~\bibnamefont {Haroche}},\ }\bibfield  {title}
	{\enquote {\bibinfo {title} {Observation of self-induced rabi oscillations in
				two-level atoms excited inside a resonant cavity: The ringing regime of
				superradiance},}\ }\href {\doibase 10.1103/PhysRevLett.51.1175} {\bibfield
		{journal} {\bibinfo  {journal} {Phys. Rev. Lett.}\ }\textbf {\bibinfo
			{volume} {51}},\ \bibinfo {pages} {1175--1178} (\bibinfo {year}
		{1983})}\BibitemShut {NoStop}%
	\bibitem [{\citenamefont {Sonnefraud}\ \emph {et~al.}(2010)\citenamefont
		{Sonnefraud}, \citenamefont {Verellen}, \citenamefont {Sobhani},
		\citenamefont {Vandenbosch}, \citenamefont {Moshchalkov}, \citenamefont
		{Van~Dorpe}, \citenamefont {Nordlander},\ and\ \citenamefont
		{Maier}}]{sonnefraud_nano_2010}%
	\BibitemOpen
	\bibfield  {author} {\bibinfo {author} {\bibfnamefont {Y.}~\bibnamefont
			{Sonnefraud}}, \bibinfo {author} {\bibfnamefont {N.}~\bibnamefont
			{Verellen}}, \bibinfo {author} {\bibfnamefont {H.}~\bibnamefont {Sobhani}},
		\bibinfo {author} {\bibfnamefont {G.~A.~E.}\ \bibnamefont {Vandenbosch}},
		\bibinfo {author} {\bibfnamefont {V.~V.}\ \bibnamefont {Moshchalkov}},
		\bibinfo {author} {\bibfnamefont {P.}~\bibnamefont {Van~Dorpe}}, \bibinfo
		{author} {\bibfnamefont {P.}~\bibnamefont {Nordlander}}, \ and\ \bibinfo
		{author} {\bibfnamefont {S.~A.}\ \bibnamefont {Maier}},\ }\bibfield  {title}
	{\enquote {\bibinfo {title} {Experimental realization of subradiant,
				superradiant, and fano resonances in ring/disk plasmonic nanocavities},}\
	}\href {\doibase 10.1021/nn901580r} {\bibfield  {journal} {\bibinfo
			{journal} {ACS Nano}\ }\textbf {\bibinfo {volume} {4}},\ \bibinfo {pages}
		{1664--1670} (\bibinfo {year} {2010})}\BibitemShut {NoStop}%
	\bibitem [{\citenamefont {Wenclawiak}\ \emph {et~al.}(2017)\citenamefont
		{Wenclawiak}, \citenamefont {Unterrainer},\ and\ \citenamefont
		{Darmo}}]{wenclawiak_apl_2017}%
	\BibitemOpen
	\bibfield  {author} {\bibinfo {author} {\bibfnamefont {M.}~\bibnamefont
			{Wenclawiak}}, \bibinfo {author} {\bibfnamefont {K.}~\bibnamefont
			{Unterrainer}}, \ and\ \bibinfo {author} {\bibfnamefont {J.}~\bibnamefont
			{Darmo}},\ }\bibfield  {title} {\enquote {\bibinfo {title} {Cooperative
				effects in an ensemble of planar meta-atoms},}\ }\href {\doibase
		10.1063/1.4989691} {\bibfield  {journal} {\bibinfo  {journal} {Appl. Phys.
				Lett.}\ }\textbf {\bibinfo {volume} {110}},\ \bibinfo {pages} {261101}
		(\bibinfo {year} {2017})},\ \Eprint
	{http://arxiv.org/abs/https://doi.org/10.1063/1.4989691}
	{https://doi.org/10.1063/1.4989691} \BibitemShut {NoStop}%
	\bibitem [{\citenamefont {Cong}\ \emph {et~al.}(2016)\citenamefont {Cong},
		\citenamefont {Zhang}, \citenamefont {Wang}, \citenamefont {Noe},
		\citenamefont {Belyanin},\ and\ \citenamefont {Kono}}]{cong_josab_2016}%
	\BibitemOpen
	\bibfield  {author} {\bibinfo {author} {\bibfnamefont {Kankan}\ \bibnamefont
			{Cong}}, \bibinfo {author} {\bibfnamefont {Qi}~\bibnamefont {Zhang}},
		\bibinfo {author} {\bibfnamefont {Yongrui}\ \bibnamefont {Wang}}, \bibinfo
		{author} {\bibfnamefont {G.~Timothy}\ \bibnamefont {Noe}}, \bibinfo {author}
		{\bibfnamefont {Alexey}\ \bibnamefont {Belyanin}}, \ and\ \bibinfo {author}
		{\bibfnamefont {Junichiro}\ \bibnamefont {Kono}},\ }\bibfield  {title}
	{\enquote {\bibinfo {title} {Dicke superradiance in solids [invited]},}\
	}\href {\doibase 10.1364/JOSAB.33.000C80} {\bibfield  {journal} {\bibinfo
			{journal} {J. Opt. Soc. Am. B}\ }\textbf {\bibinfo {volume} {33}},\ \bibinfo
		{pages} {C80--C101} (\bibinfo {year} {2016})}\BibitemShut {NoStop}%
	\bibitem [{\citenamefont {Zhang}\ \emph {et~al.}(2014)\citenamefont {Zhang},
		\citenamefont {Arikawa}, \citenamefont {Kato}, \citenamefont {Reno},
		\citenamefont {Pan}, \citenamefont {Watson}, \citenamefont {Manfra},
		\citenamefont {Zudov}, \citenamefont {Tokman}, \citenamefont {Erukhimova},
		\citenamefont {Belyanin},\ and\ \citenamefont {Kono}}]{zhang_prl_2014}%
	\BibitemOpen
	\bibfield  {author} {\bibinfo {author} {\bibfnamefont {Q.}~\bibnamefont
			{Zhang}}, \bibinfo {author} {\bibfnamefont {T.}~\bibnamefont {Arikawa}},
		\bibinfo {author} {\bibfnamefont {E.}~\bibnamefont {Kato}}, \bibinfo {author}
		{\bibfnamefont {J.~L.}\ \bibnamefont {Reno}}, \bibinfo {author}
		{\bibfnamefont {W.}~\bibnamefont {Pan}}, \bibinfo {author} {\bibfnamefont
			{J.~D.}\ \bibnamefont {Watson}}, \bibinfo {author} {\bibfnamefont {M.~J.}\
			\bibnamefont {Manfra}}, \bibinfo {author} {\bibfnamefont {M.~A.}\
			\bibnamefont {Zudov}}, \bibinfo {author} {\bibfnamefont {M.}~\bibnamefont
			{Tokman}}, \bibinfo {author} {\bibfnamefont {M.}~\bibnamefont {Erukhimova}},
		\bibinfo {author} {\bibfnamefont {A.}~\bibnamefont {Belyanin}}, \ and\
		\bibinfo {author} {\bibfnamefont {J.}~\bibnamefont {Kono}},\ }\bibfield
	{title} {\enquote {\bibinfo {title} {Superradiant decay of cyclotron
				resonance of two-dimensional electron gases},}\ }\href {\doibase
		10.1103/PhysRevLett.113.047601} {\bibfield  {journal} {\bibinfo  {journal}
			{Phys. Rev. Lett.}\ }\textbf {\bibinfo {volume} {113}},\ \bibinfo {pages}
		{047601} (\bibinfo {year} {2014})}\BibitemShut {NoStop}%
	\bibitem [{\citenamefont {Laurent}\ \emph {et~al.}(2015)\citenamefont
		{Laurent}, \citenamefont {Todorov}, \citenamefont {Vasanelli}, \citenamefont
		{Delteil}, \citenamefont {Sirtori}, \citenamefont {Sagnes},\ and\
		\citenamefont {Beaudoin}}]{laurent_prl_2015}%
	\BibitemOpen
	\bibfield  {author} {\bibinfo {author} {\bibfnamefont {T.}~\bibnamefont
			{Laurent}}, \bibinfo {author} {\bibfnamefont {Y.}~\bibnamefont {Todorov}},
		\bibinfo {author} {\bibfnamefont {A.}~\bibnamefont {Vasanelli}}, \bibinfo
		{author} {\bibfnamefont {A.}~\bibnamefont {Delteil}}, \bibinfo {author}
		{\bibfnamefont {C.}~\bibnamefont {Sirtori}}, \bibinfo {author} {\bibfnamefont
			{I.}~\bibnamefont {Sagnes}}, \ and\ \bibinfo {author} {\bibfnamefont
			{G.}~\bibnamefont {Beaudoin}},\ }\bibfield  {title} {\enquote {\bibinfo
			{title} {Superradiant emission from a collective excitation in a
				semiconductor},}\ }\href {\doibase 10.1103/PhysRevLett.115.187402} {\bibfield
		{journal} {\bibinfo  {journal} {Phys. Rev. Lett.}\ }\textbf {\bibinfo
			{volume} {115}},\ \bibinfo {pages} {187402} (\bibinfo {year}
		{2015})}\BibitemShut {NoStop}%
	\bibitem [{\citenamefont {Zhang}\ \emph {et~al.}(2016)\citenamefont {Zhang},
		\citenamefont {Lou}, \citenamefont {Li}, \citenamefont {Reno}, \citenamefont
		{Pan}, \citenamefont {Watson}, \citenamefont {Manfra},\ and\ \citenamefont
		{Kono}}]{zhang_np_2016}%
	\BibitemOpen
	\bibfield  {author} {\bibinfo {author} {\bibfnamefont {Q.}~\bibnamefont
			{Zhang}}, \bibinfo {author} {\bibfnamefont {M.}~\bibnamefont {Lou}}, \bibinfo
		{author} {\bibfnamefont {X.}~\bibnamefont {Li}}, \bibinfo {author}
		{\bibfnamefont {J.~L.}\ \bibnamefont {Reno}}, \bibinfo {author}
		{\bibfnamefont {W.}~\bibnamefont {Pan}}, \bibinfo {author} {\bibfnamefont
			{J.~D.}\ \bibnamefont {Watson}}, \bibinfo {author} {\bibfnamefont {M.~J.}\
			\bibnamefont {Manfra}}, \ and\ \bibinfo {author} {\bibfnamefont
			{J.}~\bibnamefont {Kono}},\ }\bibfield  {title} {\enquote {\bibinfo {title}
			{Collective non-perturbative coupling of 2d electrons with
				high-quality-factor terahertz cavity photons},}\ }\href {\doibase
		10.1038/NPHYS3850} {\bibfield  {journal} {\bibinfo  {journal} {Nat. Phys.}\
		}\textbf {\bibinfo {volume} {12}},\ \bibinfo {pages} {1005+} (\bibinfo {year}
		{2016})}\BibitemShut {NoStop}%
	\bibitem [{\citenamefont {Curtis}\ \emph {et~al.}(2016)\citenamefont {Curtis},
		\citenamefont {Tokumoto}, \citenamefont {Hatke}, \citenamefont {Cherian},
		\citenamefont {Reno}, \citenamefont {McGill}, \citenamefont {Karaiskaj},\
		and\ \citenamefont {Hilton}}]{curtis_prb_2016}%
	\BibitemOpen
	\bibfield  {author} {\bibinfo {author} {\bibfnamefont {J.~A.}\ \bibnamefont
			{Curtis}}, \bibinfo {author} {\bibfnamefont {T.}~\bibnamefont {Tokumoto}},
		\bibinfo {author} {\bibfnamefont {A.~T.}\ \bibnamefont {Hatke}}, \bibinfo
		{author} {\bibfnamefont {J.~G.}\ \bibnamefont {Cherian}}, \bibinfo {author}
		{\bibfnamefont {J.~L.}\ \bibnamefont {Reno}}, \bibinfo {author}
		{\bibfnamefont {S.~A.}\ \bibnamefont {McGill}}, \bibinfo {author}
		{\bibfnamefont {D.}~\bibnamefont {Karaiskaj}}, \ and\ \bibinfo {author}
		{\bibfnamefont {D.~J.}\ \bibnamefont {Hilton}},\ }\bibfield  {title}
	{\enquote {\bibinfo {title} {Cyclotron decay time of a two-dimensional
				electron gas from 0.4 to 100 {K}},}\ }\href {\doibase
		10.1103/PhysRevB.93.155437} {\bibfield  {journal} {\bibinfo  {journal} {Phys.
				Rev. B}\ }\textbf {\bibinfo {volume} {93}},\ \bibinfo {pages} {155437}
		(\bibinfo {year} {2016})}\BibitemShut {NoStop}%
	\bibitem [{\citenamefont {Maag}\ \emph {et~al.}(2016)\citenamefont {Maag},
		\citenamefont {Bayer}, \citenamefont {Baierl}, \citenamefont {Hohenleutner},
		\citenamefont {Korn}, \citenamefont {Schueller}, \citenamefont {Schuh},
		\citenamefont {Bougeard}, \citenamefont {Lange}, \citenamefont {Huber},
		\citenamefont {Mootz}, \citenamefont {Sipe}, \citenamefont {Koch},\ and\
		\citenamefont {Kira}}]{maag_np_2016}%
	\BibitemOpen
	\bibfield  {author} {\bibinfo {author} {\bibfnamefont {T.}~\bibnamefont
			{Maag}}, \bibinfo {author} {\bibfnamefont {A.}~\bibnamefont {Bayer}},
		\bibinfo {author} {\bibfnamefont {S.}~\bibnamefont {Baierl}}, \bibinfo
		{author} {\bibfnamefont {M.}~\bibnamefont {Hohenleutner}}, \bibinfo {author}
		{\bibfnamefont {T.}~\bibnamefont {Korn}}, \bibinfo {author} {\bibfnamefont
			{C.}~\bibnamefont {Schueller}}, \bibinfo {author} {\bibfnamefont
			{D.}~\bibnamefont {Schuh}}, \bibinfo {author} {\bibfnamefont
			{D.}~\bibnamefont {Bougeard}}, \bibinfo {author} {\bibfnamefont
			{C.}~\bibnamefont {Lange}}, \bibinfo {author} {\bibfnamefont
			{R.}~\bibnamefont {Huber}}, \bibinfo {author} {\bibfnamefont
			{M.}~\bibnamefont {Mootz}}, \bibinfo {author} {\bibfnamefont {J.~E.}\
			\bibnamefont {Sipe}}, \bibinfo {author} {\bibfnamefont {S.~W.}\ \bibnamefont
			{Koch}}, \ and\ \bibinfo {author} {\bibfnamefont {M.}~\bibnamefont {Kira}},\
	}\bibfield  {title} {\enquote {\bibinfo {title} {Coherent cyclotron motion
				beyond {K}ohn's theorem},}\ }\href {\doibase 10.1038/NPHYS3559} {\bibfield
		{journal} {\bibinfo  {journal} {Nat. Phys.}\ }\textbf {\bibinfo {volume}
			{12}},\ \bibinfo {pages} {119} (\bibinfo {year} {2016})}\BibitemShut
	{NoStop}%
	\bibitem [{\citenamefont {Muravev}\ \emph {et~al.}(2017)\citenamefont
		{Muravev}, \citenamefont {Andreev}, \citenamefont {Belyanin}, \citenamefont
		{Gubarev},\ and\ \citenamefont {Kukushkin}}]{muravev_prb_2017}%
	\BibitemOpen
	\bibfield  {author} {\bibinfo {author} {\bibfnamefont {V.~M.}\ \bibnamefont
			{Muravev}}, \bibinfo {author} {\bibfnamefont {I.~V.}\ \bibnamefont
			{Andreev}}, \bibinfo {author} {\bibfnamefont {V.~N.}\ \bibnamefont
			{Belyanin}}, \bibinfo {author} {\bibfnamefont {S.~I.}\ \bibnamefont
			{Gubarev}}, \ and\ \bibinfo {author} {\bibfnamefont {I.~V.}\ \bibnamefont
			{Kukushkin}},\ }\bibfield  {title} {\enquote {\bibinfo {title} {Observation
				of axisymmetric dark plasma excitations in a two-dimensional electron
				system},}\ }\href {\doibase 10.1103/PhysRevB.96.045421} {\bibfield  {journal}
		{\bibinfo  {journal} {Phys. Rev. B}\ }\textbf {\bibinfo {volume} {96}},\
		\bibinfo {pages} {045421} (\bibinfo {year} {2017})}\BibitemShut {NoStop}%
	\bibitem [{\citenamefont {Moiseev}\ and\ \citenamefont
		{Greenberg}(2017)}]{moiseev_prb_2017}%
	\BibitemOpen
	\bibfield  {author} {\bibinfo {author} {\bibfnamefont {A.~G.}\ \bibnamefont
			{Moiseev}}\ and\ \bibinfo {author} {\bibfnamefont {Ya.~S.}\ \bibnamefont
			{Greenberg}},\ }\bibfield  {title} {\enquote {\bibinfo {title} {Single-photon
				superradiant decay of cyclotron resonance in a $p$-type single-crystal
				semiconductor film with cubic structure},}\ }\href {\doibase
		10.1103/PhysRevB.96.075208} {\bibfield  {journal} {\bibinfo  {journal} {Phys.
				Rev. B}\ }\textbf {\bibinfo {volume} {96}},\ \bibinfo {pages} {075208}
		(\bibinfo {year} {2017})}\BibitemShut {NoStop}%
	\bibitem [{\citenamefont {Jackson}(1999)}]{jackson_book}%
	\BibitemOpen
	\bibfield  {author} {\bibinfo {author} {\bibfnamefont {J.~D.}\ \bibnamefont
			{Jackson}},\ }\href@noop {} {\emph {\bibinfo {title} {Classical
				electrodynamics}}},\ \bibinfo {edition} {3rd}\ ed.\ (\bibinfo  {publisher}
	{Wiley},\ \bibinfo {address} {New York, NY},\ \bibinfo {year}
	{1999})\BibitemShut {NoStop}%
	\bibitem [{\citenamefont {Nuss}\ and\ \citenamefont
		{Orenstein}(1998)}]{nuss_book}%
	\BibitemOpen
	\bibfield  {author} {\bibinfo {author} {\bibfnamefont {M.C.}\ \bibnamefont
			{Nuss}}\ and\ \bibinfo {author} {\bibfnamefont {J.}~\bibnamefont
			{Orenstein}},\ }\bibfield  {title} {\enquote {\bibinfo {title} {Terahertz
				time-domain spectroscopy},}\ }in\ \href@noop {} {\emph {\bibinfo {booktitle}
			{Millimeter And Submillimeter Wave Spectroscopy Of Solids}}},\ \bibinfo
	{series} {Topics in Applied Physics}, Vol.~\bibinfo {volume} {74},\ \bibinfo
	{editor} {edited by\ \bibinfo {editor} {\bibfnamefont {G.}~\bibnamefont
			{Gruner}}}\ (\bibinfo  {publisher} {Springer, Berlin},\ \bibinfo {year}
	{1998})\ pp.\ \bibinfo {pages} {7--50}\BibitemShut {NoStop}%
	\bibitem [{\citenamefont {Kozlov}\ and\ \citenamefont
		{Volkov}(1998)}]{kozlov_book}%
	\BibitemOpen
	\bibfield  {author} {\bibinfo {author} {\bibfnamefont {G.~V.}\ \bibnamefont
			{Kozlov}}\ and\ \bibinfo {author} {\bibfnamefont {A.~A.}\ \bibnamefont
			{Volkov}},\ }\bibfield  {title} {\enquote {\bibinfo {title} {Coherent source
				submillimeter wave spectroscopy},}\ }in\ \href@noop {} {\emph {\bibinfo
			{booktitle} {Millimeter and Submillimeter Wave Spectroscopy of Solids}}},\
	\bibinfo {editor} {edited by\ \bibinfo {editor} {\bibfnamefont
			{G.}~\bibnamefont {Gr\"{u}ner}}}\ (\bibinfo  {publisher} {Springer},\
	\bibinfo {address} {Berlin},\ \bibinfo {year} {1998})\ p.~\bibinfo {pages}
	{51}\BibitemShut {NoStop}%
	\bibitem [{\citenamefont {Hasan}\ and\ \citenamefont
		{Kane}(2010)}]{hasan_rmp_2010}%
	\BibitemOpen
	\bibfield  {author} {\bibinfo {author} {\bibfnamefont {M.~Z.}\ \bibnamefont
			{Hasan}}\ and\ \bibinfo {author} {\bibfnamefont {C.~L.}\ \bibnamefont
			{Kane}},\ }\bibfield  {title} {\enquote {\bibinfo {title}
			{\textit{Colloquium}: Topological insulators},}\ }\href {\doibase
		10.1103/RevModPhys.82.3045} {\bibfield  {journal} {\bibinfo  {journal} {Rev.
				Mod. Phys.}\ }\textbf {\bibinfo {volume} {82}},\ \bibinfo {pages}
		{3045--3067} (\bibinfo {year} {2010})}\BibitemShut {NoStop}%
	\bibitem [{\citenamefont {Qi}\ and\ \citenamefont {Zhang}(2011)}]{qi_rmp_2011}%
	\BibitemOpen
	\bibfield  {author} {\bibinfo {author} {\bibfnamefont {Xiao-Liang}\
			\bibnamefont {Qi}}\ and\ \bibinfo {author} {\bibfnamefont {Shou-Cheng}\
			\bibnamefont {Zhang}},\ }\bibfield  {title} {\enquote {\bibinfo {title}
			{Topological insulators and superconductors},}\ }\href {\doibase
		10.1103/RevModPhys.83.1057} {\bibfield  {journal} {\bibinfo  {journal} {Rev.
				Mod. Phys.}\ }\textbf {\bibinfo {volume} {83}},\ \bibinfo {pages}
		{1057--1110} (\bibinfo {year} {2011})}\BibitemShut {NoStop}%
	\bibitem [{\citenamefont {Br\"une}\ \emph {et~al.}(2011)\citenamefont
		{Br\"une}, \citenamefont {Liu}, \citenamefont {Novik}, \citenamefont
		{Hankiewicz}, \citenamefont {Buhmann}, \citenamefont {Chen}, \citenamefont
		{Qi}, \citenamefont {Shen}, \citenamefont {Zhang},\ and\ \citenamefont
		{Molenkamp}}]{brune_prl_2011}%
	\BibitemOpen
	\bibfield  {author} {\bibinfo {author} {\bibfnamefont {C.}~\bibnamefont
			{Br\"une}}, \bibinfo {author} {\bibfnamefont {C.~X.}\ \bibnamefont {Liu}},
		\bibinfo {author} {\bibfnamefont {E.~G.}\ \bibnamefont {Novik}}, \bibinfo
		{author} {\bibfnamefont {E.~M.}\ \bibnamefont {Hankiewicz}}, \bibinfo
		{author} {\bibfnamefont {H.}~\bibnamefont {Buhmann}}, \bibinfo {author}
		{\bibfnamefont {Y.~L.}\ \bibnamefont {Chen}}, \bibinfo {author}
		{\bibfnamefont {X.~L.}\ \bibnamefont {Qi}}, \bibinfo {author} {\bibfnamefont
			{Z.~X.}\ \bibnamefont {Shen}}, \bibinfo {author} {\bibfnamefont {S.~C.}\
			\bibnamefont {Zhang}}, \ and\ \bibinfo {author} {\bibfnamefont {L.~W.}\
			\bibnamefont {Molenkamp}},\ }\bibfield  {title} {\enquote {\bibinfo {title}
			{Quantum {H}all effect from the topological surface states of strained bulk
				$\mathrm{HgTe}$},}\ }\href {\doibase 10.1103/PhysRevLett.106.126803}
	{\bibfield  {journal} {\bibinfo  {journal} {Phys. Rev. Lett.}\ }\textbf
		{\bibinfo {volume} {106}},\ \bibinfo {pages} {126803} (\bibinfo {year}
		{2011})}\BibitemShut {NoStop}%
	\bibitem [{\citenamefont {Kozlov}\ \emph {et~al.}(2014)\citenamefont {Kozlov},
		\citenamefont {Kvon}, \citenamefont {Olshanetsky}, \citenamefont {Mikhailov},
		\citenamefont {Dvoretsky},\ and\ \citenamefont {Weiss}}]{Kozlov2014}%
	\BibitemOpen
	\bibfield  {author} {\bibinfo {author} {\bibfnamefont
			{D.{\hspace{0.167em}}A.}\ \bibnamefont {Kozlov}}, \bibinfo {author}
		{\bibfnamefont {Z.{\hspace{0.167em}}D.}\ \bibnamefont {Kvon}}, \bibinfo
		{author} {\bibfnamefont {E.{\hspace{0.167em}}B.}\ \bibnamefont
			{Olshanetsky}}, \bibinfo {author} {\bibfnamefont {N.{\hspace{0.167em}}N.}\
			\bibnamefont {Mikhailov}}, \bibinfo {author} {\bibfnamefont
			{S.{\hspace{0.167em}}A.}\ \bibnamefont {Dvoretsky}}, \ and\ \bibinfo {author}
		{\bibfnamefont {D.}~\bibnamefont {Weiss}},\ }\bibfield  {title} {\enquote
		{\bibinfo {title} {Transport properties of a 3d topological insulator based
				on a strained high-mobility {HgTe} film},}\ }\href {\doibase
		10.1103/physrevlett.112.196801} {\bibfield  {journal} {\bibinfo  {journal}
			{Physical Review Letters}\ }\textbf {\bibinfo {volume} {112}} (\bibinfo
		{year} {2014}),\ 10.1103/physrevlett.112.196801}\BibitemShut {NoStop}%
	\bibitem [{\citenamefont {Shuvaev}\ \emph
		{et~al.}(2013{\natexlab{a}})\citenamefont {Shuvaev}, \citenamefont
		{Astakhov}, \citenamefont {Tkachov}, \citenamefont {Br\"une}, \citenamefont
		{Buhmann}, \citenamefont {Molenkamp},\ and\ \citenamefont
		{Pimenov}}]{shuvaev_prb_2013}%
	\BibitemOpen
	\bibfield  {author} {\bibinfo {author} {\bibfnamefont {A.~M.}\ \bibnamefont
			{Shuvaev}}, \bibinfo {author} {\bibfnamefont {G.~V.}\ \bibnamefont
			{Astakhov}}, \bibinfo {author} {\bibfnamefont {G.}~\bibnamefont {Tkachov}},
		\bibinfo {author} {\bibfnamefont {C.}~\bibnamefont {Br\"une}}, \bibinfo
		{author} {\bibfnamefont {H.}~\bibnamefont {Buhmann}}, \bibinfo {author}
		{\bibfnamefont {L.~W.}\ \bibnamefont {Molenkamp}}, \ and\ \bibinfo {author}
		{\bibfnamefont {A.}~\bibnamefont {Pimenov}},\ }\bibfield  {title} {\enquote
		{\bibinfo {title} {Terahertz quantum {H}all effect of {D}irac fermions in a
				topological insulator},}\ }\href {\doibase 10.1103/PhysRevB.87.121104}
	{\bibfield  {journal} {\bibinfo  {journal} {Phys. Rev. B}\ }\textbf {\bibinfo
			{volume} {87}},\ \bibinfo {pages} {121104} (\bibinfo {year}
		{2013}{\natexlab{a}})}\BibitemShut {NoStop}%
	\bibitem [{\citenamefont {Dziom}\ \emph
		{et~al.}(2017{\natexlab{a}})\citenamefont {Dziom}, \citenamefont {Shuvaev},
		\citenamefont {Pimenov}, \citenamefont {Astakhov}, \citenamefont {Ames},
		\citenamefont {Bendias}, \citenamefont {B{\"o}ttcher}, \citenamefont
		{Tkachov}, \citenamefont {Hankiewicz}, \citenamefont {Br{\"u}ne},
		\citenamefont {Buhmann},\ and\ \citenamefont {Molenkamp}}]{dziom_ncomm_2017}%
	\BibitemOpen
	\bibfield  {author} {\bibinfo {author} {\bibfnamefont {V.}~\bibnamefont
			{Dziom}}, \bibinfo {author} {\bibfnamefont {A.}~\bibnamefont {Shuvaev}},
		\bibinfo {author} {\bibfnamefont {A.}~\bibnamefont {Pimenov}}, \bibinfo
		{author} {\bibfnamefont {G.~V.}\ \bibnamefont {Astakhov}}, \bibinfo {author}
		{\bibfnamefont {C.}~\bibnamefont {Ames}}, \bibinfo {author} {\bibfnamefont
			{K.}~\bibnamefont {Bendias}}, \bibinfo {author} {\bibfnamefont
			{J.}~\bibnamefont {B{\"o}ttcher}}, \bibinfo {author} {\bibfnamefont
			{G.}~\bibnamefont {Tkachov}}, \bibinfo {author} {\bibfnamefont {E.~M.}\
			\bibnamefont {Hankiewicz}}, \bibinfo {author} {\bibfnamefont
			{C.}~\bibnamefont {Br{\"u}ne}}, \bibinfo {author} {\bibfnamefont
			{H.}~\bibnamefont {Buhmann}}, \ and\ \bibinfo {author} {\bibfnamefont
			{L.~W.}\ \bibnamefont {Molenkamp}},\ }\bibfield  {title} {\enquote {\bibinfo
			{title} {Observation of the universal magnetoelectric effect in a 3{D}
				topological insulator},}\ }\href {http://dx.doi.org/10.1038/ncomms15197}
	{\bibfield  {journal} {\bibinfo  {journal} {Nat. Commun.}\ }\textbf {\bibinfo
			{volume} {8}},\ \bibinfo {pages} {15197} (\bibinfo {year}
		{2017}{\natexlab{a}})}\BibitemShut {NoStop}%
	\bibitem [{\citenamefont {Kvon}\ \emph {et~al.}(2009)\citenamefont {Kvon},
		\citenamefont {Olshanetsky}, \citenamefont {Mikhailov},\ and\ \citenamefont
		{Kozlov}}]{kvon_ltp_2009}%
	\BibitemOpen
	\bibfield  {author} {\bibinfo {author} {\bibfnamefont {Z.~D.}\ \bibnamefont
			{Kvon}}, \bibinfo {author} {\bibfnamefont {E.~B.}\ \bibnamefont
			{Olshanetsky}}, \bibinfo {author} {\bibfnamefont {N.~N.}\ \bibnamefont
			{Mikhailov}}, \ and\ \bibinfo {author} {\bibfnamefont {D.~A.}\ \bibnamefont
			{Kozlov}},\ }\bibfield  {title} {\enquote {\bibinfo {title} {Two-dimensional
				electron systems in $\mathrm{HgTe}$ quantum wells},}\ }\href {\doibase
		10.1063/1.3064862} {\bibfield  {journal} {\bibinfo  {journal} {Low Temp.
				Phys.}\ }\textbf {\bibinfo {volume} {35}},\ \bibinfo {pages} {6--14}
		(\bibinfo {year} {2009})}\BibitemShut {NoStop}%
	\bibitem [{\citenamefont {Shuvaev}\ \emph
		{et~al.}(2013{\natexlab{b}})\citenamefont {Shuvaev}, \citenamefont {Pimenov},
		\citenamefont {Astakhov}, \citenamefont {Muhlbauer}, \citenamefont {Brune},
		\citenamefont {Buhmann},\ and\ \citenamefont {Molenkamp}}]{shuvaev_apl_2013}%
	\BibitemOpen
	\bibfield  {author} {\bibinfo {author} {\bibfnamefont {A.}~\bibnamefont
			{Shuvaev}}, \bibinfo {author} {\bibfnamefont {A.}~\bibnamefont {Pimenov}},
		\bibinfo {author} {\bibfnamefont {G.~V.}\ \bibnamefont {Astakhov}}, \bibinfo
		{author} {\bibfnamefont {M.}~\bibnamefont {Muhlbauer}}, \bibinfo {author}
		{\bibfnamefont {C.}~\bibnamefont {Brune}}, \bibinfo {author} {\bibfnamefont
			{H.}~\bibnamefont {Buhmann}}, \ and\ \bibinfo {author} {\bibfnamefont
			{L.~W.}\ \bibnamefont {Molenkamp}},\ }\bibfield  {title} {\enquote {\bibinfo
			{title} {Room temperature electrically tunable terahertz {F}araday effect},}\
	}\href {\doibase 10.1063/1.4811496} {\bibfield  {journal} {\bibinfo
			{journal} {Appl. Phys. Lett.}\ }\textbf {\bibinfo {volume} {102}},\ \bibinfo
		{eid} {241902} (\bibinfo {year} {2013}{\natexlab{b}})}\BibitemShut {NoStop}%
	\bibitem [{\citenamefont {Kozlov}\ \emph {et~al.}(2016)\citenamefont {Kozlov},
		\citenamefont {Bauer}, \citenamefont {Ziegler}, \citenamefont {Fischer},
		\citenamefont {Savchenko}, \citenamefont {Kvon}, \citenamefont {Mikhailov},
		\citenamefont {Dvoretsky},\ and\ \citenamefont {Weiss}}]{kozlov_prl_2016}%
	\BibitemOpen
	\bibfield  {author} {\bibinfo {author} {\bibfnamefont {D.~A.}\ \bibnamefont
			{Kozlov}}, \bibinfo {author} {\bibfnamefont {D.}~\bibnamefont {Bauer}},
		\bibinfo {author} {\bibfnamefont {J.}~\bibnamefont {Ziegler}}, \bibinfo
		{author} {\bibfnamefont {R.}~\bibnamefont {Fischer}}, \bibinfo {author}
		{\bibfnamefont {M.~L.}\ \bibnamefont {Savchenko}}, \bibinfo {author}
		{\bibfnamefont {Z.~D.}\ \bibnamefont {Kvon}}, \bibinfo {author}
		{\bibfnamefont {N.~N.}\ \bibnamefont {Mikhailov}}, \bibinfo {author}
		{\bibfnamefont {S.~A.}\ \bibnamefont {Dvoretsky}}, \ and\ \bibinfo {author}
		{\bibfnamefont {D.}~\bibnamefont {Weiss}},\ }\bibfield  {title} {\enquote
		{\bibinfo {title} {Probing quantum capacitance in a 3{D} topological
				insulator},}\ }\href {\doibase 10.1103/PhysRevLett.116.166802} {\bibfield
		{journal} {\bibinfo  {journal} {Phys. Rev. Lett.}\ }\textbf {\bibinfo
			{volume} {116}},\ \bibinfo {pages} {166802} (\bibinfo {year}
		{2016})}\BibitemShut {NoStop}%
	\bibitem [{\citenamefont {Volkov}\ \emph {et~al.}(1985)\citenamefont {Volkov},
		\citenamefont {Goncharov}, \citenamefont {Kozlov}, \citenamefont {Lebedev},\
		and\ \citenamefont {Prokhorov}}]{volkov_infrared_1985}%
	\BibitemOpen
	\bibfield  {author} {\bibinfo {author} {\bibfnamefont {A.~A.}\ \bibnamefont
			{Volkov}}, \bibinfo {author} {\bibfnamefont {Yu.~G.}\ \bibnamefont
			{Goncharov}}, \bibinfo {author} {\bibfnamefont {G.~V.}\ \bibnamefont
			{Kozlov}}, \bibinfo {author} {\bibfnamefont {S.~P.}\ \bibnamefont {Lebedev}},
		\ and\ \bibinfo {author} {\bibfnamefont {A.~M.}\ \bibnamefont {Prokhorov}},\
	}\bibfield  {title} {\enquote {\bibinfo {title} {Dielectric measurements in
				the submillimeter wavelength region},}\ }\href {\doibase DOI:
		10.1016/0020-0891(85)90109-5} {\bibfield  {journal} {\bibinfo  {journal}
			{Infrared Phys.}\ }\textbf {\bibinfo {volume} {25}},\ \bibinfo {pages} {369}
		(\bibinfo {year} {1985})}\BibitemShut {NoStop}%
	\bibitem [{\citenamefont {Shuvaev}\ \emph {et~al.}(2012)\citenamefont
		{Shuvaev}, \citenamefont {Astakhov}, \citenamefont {Br\"{u}ne}, \citenamefont
		{Buhmann}, \citenamefont {Molenkamp},\ and\ \citenamefont
		{Pimenov}}]{shuvaev_sst_2012}%
	\BibitemOpen
	\bibfield  {author} {\bibinfo {author} {\bibfnamefont {A.~M.}\ \bibnamefont
			{Shuvaev}}, \bibinfo {author} {\bibfnamefont {G.~V.}\ \bibnamefont
			{Astakhov}}, \bibinfo {author} {\bibfnamefont {C.}~\bibnamefont {Br\"{u}ne}},
		\bibinfo {author} {\bibfnamefont {H.}~\bibnamefont {Buhmann}}, \bibinfo
		{author} {\bibfnamefont {L.~W.}\ \bibnamefont {Molenkamp}}, \ and\ \bibinfo
		{author} {\bibfnamefont {A.}~\bibnamefont {Pimenov}},\ }\bibfield  {title}
	{\enquote {\bibinfo {title} {Terahertz magneto-optical spectroscopy in
				{$\mathrm{HgTe}$} thin films},}\ }\href
	{http://stacks.iop.org/0268-1242/27/i=12/a=124004} {\bibfield  {journal}
		{\bibinfo  {journal} {Semicond. Sci. Technol.}\ }\textbf {\bibinfo {volume}
			{27}},\ \bibinfo {pages} {124004} (\bibinfo {year} {2012})}\BibitemShut
	{NoStop}%
	\bibitem [{\citenamefont {Matov}\ \emph {et~al.}(1996)\citenamefont {Matov},
		\citenamefont {Meshkov}, \citenamefont {Polishchuk},\ and\ \citenamefont
		{Popov}}]{matov_jetp_1996}%
	\BibitemOpen
	\bibfield  {author} {\bibinfo {author} {\bibfnamefont {O.~R.}\ \bibnamefont
			{Matov}}, \bibinfo {author} {\bibfnamefont {O.~F.}\ \bibnamefont {Meshkov}},
		\bibinfo {author} {\bibfnamefont {O.~V.}\ \bibnamefont {Polishchuk}}, \ and\
		\bibinfo {author} {\bibfnamefont {V.~V.}\ \bibnamefont {Popov}},\ }\bibfield
	{title} {\enquote {\bibinfo {title} {Theory of electromagnetic emission of
				2-dimensional magnetoplasma and cyclotron oscillations in semiconducting
				heterostructure with periodic screen},}\ }\href@noop {} {\bibfield  {journal}
		{\bibinfo  {journal} {J. Exp. Theor. Phys.}\ }\textbf {\bibinfo {volume}
			{109}},\ \bibinfo {pages} {876--890} (\bibinfo {year} {1996})}\BibitemShut
	{NoStop}%
	\bibitem [{\citenamefont {Mikhailov}(1996)}]{mikhailov_prb_1996}%
	\BibitemOpen
	\bibfield  {author} {\bibinfo {author} {\bibfnamefont {S.~A.}\ \bibnamefont
			{Mikhailov}},\ }\bibfield  {title} {\enquote {\bibinfo {title} {Radiative
				decay of collective excitations in an array of quantum dots},}\ }\href
	{\doibase 10.1103/PhysRevB.54.10335} {\bibfield  {journal} {\bibinfo
			{journal} {Phys. Rev. B}\ }\textbf {\bibinfo {volume} {54}},\ \bibinfo
		{pages} {10335--10338} (\bibinfo {year} {1996})}\BibitemShut {NoStop}%
	\bibitem [{\citenamefont {Mikhailov}(2004)}]{mikhailov_prb_2004}%
	\BibitemOpen
	\bibfield  {author} {\bibinfo {author} {\bibfnamefont {S.~A.}\ \bibnamefont
			{Mikhailov}},\ }\bibfield  {title} {\enquote {\bibinfo {title}
			{Microwave-induced magnetotransport phenomena in two-dimensional electron
				systems: Importance of electrodynamic effects},}\ }\href {\doibase
		10.1103/PhysRevB.70.165311} {\bibfield  {journal} {\bibinfo  {journal} {Phys.
				Rev. B}\ }\textbf {\bibinfo {volume} {70}},\ \bibinfo {pages} {165311}
		(\bibinfo {year} {2004})}\BibitemShut {NoStop}%
	\bibitem [{\citenamefont {Ashcroft}\ and\ \citenamefont
		{Mermin}(1976)}]{ashcroft_book}%
	\BibitemOpen
	\bibfield  {author} {\bibinfo {author} {\bibfnamefont {N.~W.}\ \bibnamefont
			{Ashcroft}}\ and\ \bibinfo {author} {\bibfnamefont {N.~D.}\ \bibnamefont
			{Mermin}},\ }\href@noop {} {\emph {\bibinfo {title} {Solid State Physics}}},\
	HRW international editions\ (\bibinfo  {publisher} {Holt, Rinehart and
		Winston},\ \bibinfo {year} {1976})\BibitemShut {NoStop}%
	\bibitem [{\citenamefont {Shuvaev}\ \emph {et~al.}(2017)\citenamefont
		{Shuvaev}, \citenamefont {Dziom}, \citenamefont {Mikhailov}, \citenamefont
		{Kvon}, \citenamefont {Shao}, \citenamefont {Basov},\ and\ \citenamefont
		{Pimenov}}]{shuvaev_prb_2017}%
	\BibitemOpen
	\bibfield  {author} {\bibinfo {author} {\bibfnamefont {A.~M.}\ \bibnamefont
			{Shuvaev}}, \bibinfo {author} {\bibfnamefont {V.}~\bibnamefont {Dziom}},
		\bibinfo {author} {\bibfnamefont {N.~N.}\ \bibnamefont {Mikhailov}}, \bibinfo
		{author} {\bibfnamefont {Z.~D.}\ \bibnamefont {Kvon}}, \bibinfo {author}
		{\bibfnamefont {Y.}~\bibnamefont {Shao}}, \bibinfo {author} {\bibfnamefont
			{D.~N.}\ \bibnamefont {Basov}}, \ and\ \bibinfo {author} {\bibfnamefont
			{A.}~\bibnamefont {Pimenov}},\ }\bibfield  {title} {\enquote {\bibinfo
			{title} {Band structure of a two-dimensional {D}irac semimetal from cyclotron
				resonance},}\ }\href {\doibase 10.1103/PhysRevB.96.155434} {\bibfield
		{journal} {\bibinfo  {journal} {Phys. Rev. B}\ }\textbf {\bibinfo {volume}
			{96}},\ \bibinfo {pages} {155434} (\bibinfo {year} {2017})}\BibitemShut
	{NoStop}%
	\bibitem [{\citenamefont {Dziom}\ \emph
		{et~al.}(2017{\natexlab{b}})\citenamefont {Dziom}, \citenamefont {Shuvaev},
		\citenamefont {Mikhailov},\ and\ \citenamefont {Pimenov}}]{dziom_2d_2017}%
	\BibitemOpen
	\bibfield  {author} {\bibinfo {author} {\bibfnamefont {V.}~\bibnamefont
			{Dziom}}, \bibinfo {author} {\bibfnamefont {A.}~\bibnamefont {Shuvaev}},
		\bibinfo {author} {\bibfnamefont {N.~N.}\ \bibnamefont {Mikhailov}}, \ and\
		\bibinfo {author} {\bibfnamefont {A.}~\bibnamefont {Pimenov}},\ }\bibfield
	{title} {\enquote {\bibinfo {title} {Terahertz properties of {D}irac fermions
				in $\mathrm{HgTe}$ films with optical doping},}\ }\href
	{http://stacks.iop.org/2053-1583/4/i=2/a=024005} {\bibfield  {journal}
		{\bibinfo  {journal} {2D Mater.}\ }\textbf {\bibinfo {volume} {4}},\ \bibinfo
		{pages} {024005} (\bibinfo {year} {2017}{\natexlab{b}})}\BibitemShut
	{NoStop}%
	\bibitem [{\citenamefont {Dressel}\ and\ \citenamefont
		{Gr\"{u}ner}(2002)}]{dressel_book_2002}%
	\BibitemOpen
	\bibfield  {author} {\bibinfo {author} {\bibfnamefont {M.}~\bibnamefont
			{Dressel}}\ and\ \bibinfo {author} {\bibfnamefont {G.}~\bibnamefont
			{Gr\"{u}ner}},\ }\href@noop {} {\emph {\bibinfo {title} {Electrodynamics of
				Solids: Optical Properties of Electrons in Matter}}},\ \bibinfo {edition}
	{1st}\ ed.\ (\bibinfo  {publisher} {Cambridge University Press},\ \bibinfo
	{address} {Cambridge},\ \bibinfo {year} {2002})\ p.~\bibinfo {pages}
	{67}\BibitemShut {NoStop}%
	\bibitem [{\citenamefont {Shuvaev}\ \emph {et~al.}(2016)\citenamefont
		{Shuvaev}, \citenamefont {Dziom}, \citenamefont {Kvon}, \citenamefont
		{Mikhailov},\ and\ \citenamefont {Pimenov}}]{shuvaev_prl_2016}%
	\BibitemOpen
	\bibfield  {author} {\bibinfo {author} {\bibfnamefont {A.}~\bibnamefont
			{Shuvaev}}, \bibinfo {author} {\bibfnamefont {V.}~\bibnamefont {Dziom}},
		\bibinfo {author} {\bibfnamefont {Z.~D.}\ \bibnamefont {Kvon}}, \bibinfo
		{author} {\bibfnamefont {N.~N.}\ \bibnamefont {Mikhailov}}, \ and\ \bibinfo
		{author} {\bibfnamefont {A.}~\bibnamefont {Pimenov}},\ }\bibfield  {title}
	{\enquote {\bibinfo {title} {Universal {F}araday rotation in $\mathrm{HgTe}$
				wells with critical thickness},}\ }\href {\doibase
		10.1103/PhysRevLett.117.117401} {\bibfield  {journal} {\bibinfo  {journal}
			{Phys. Rev. Lett.}\ }\textbf {\bibinfo {volume} {117}},\ \bibinfo {pages}
		{117401} (\bibinfo {year} {2016})}\BibitemShut {NoStop}%
	\bibitem [{\citenamefont {Berchenko}\ and\ \citenamefont
		{Pashkovski{\u{\i}}}(1976)}]{Berchenko1976}%
	\BibitemOpen
	\bibfield  {author} {\bibinfo {author} {\bibfnamefont {N~N}\ \bibnamefont
			{Berchenko}}\ and\ \bibinfo {author} {\bibfnamefont {M~V}\ \bibnamefont
			{Pashkovski{\u{\i}}}},\ }\bibfield  {title} {\enquote {\bibinfo {title}
			{Mercury telluride{\textemdash}a zero-gap semiconductor},}\ }\href {\doibase
		10.1070/pu1976v019n06abeh005265} {\bibfield  {journal} {\bibinfo  {journal}
			{Soviet Physics Uspekhi}\ }\textbf {\bibinfo {volume} {19}},\ \bibinfo
		{pages} {462--480} (\bibinfo {year} {1976})}\BibitemShut {NoStop}%
	\bibitem [{\citenamefont {Tkachov}\ \emph {et~al.}(2011)\citenamefont
		{Tkachov}, \citenamefont {Thienel}, \citenamefont {Pinneker}, \citenamefont
		{Büttner}, \citenamefont {Brüne}, \citenamefont {Buhmann}, \citenamefont
		{Molenkamp},\ and\ \citenamefont {Hankiewicz}}]{Tkachov2011}%
	\BibitemOpen
	\bibfield  {author} {\bibinfo {author} {\bibfnamefont {G.}~\bibnamefont
			{Tkachov}}, \bibinfo {author} {\bibfnamefont {C.}~\bibnamefont {Thienel}},
		\bibinfo {author} {\bibfnamefont {V.}~\bibnamefont {Pinneker}}, \bibinfo
		{author} {\bibfnamefont {B.}~\bibnamefont {Büttner}}, \bibinfo {author}
		{\bibfnamefont {C.}~\bibnamefont {Brüne}}, \bibinfo {author} {\bibfnamefont
			{H.}~\bibnamefont {Buhmann}}, \bibinfo {author} {\bibfnamefont {L.~W.}\
			\bibnamefont {Molenkamp}}, \ and\ \bibinfo {author} {\bibfnamefont {E.~M.}\
			\bibnamefont {Hankiewicz}},\ }\bibfield  {title} {\enquote {\bibinfo {title}
			{Backscattering of dirac fermions in {HgTe} quantum wells with a finite
				gap},}\ }\href {\doibase 10.1103/physrevlett.106.076802} {\bibfield
		{journal} {\bibinfo  {journal} {Physical Review Letters}\ }\textbf {\bibinfo
			{volume} {106}} (\bibinfo {year} {2011}),\
		10.1103/physrevlett.106.076802}\BibitemShut {NoStop}%
	\bibitem [{\citenamefont {Savchenko}\ \emph {et~al.}(2018)\citenamefont
		{Savchenko}, \citenamefont {Kvon}, \citenamefont {Candussio}, \citenamefont
		{Mikhailov}, \citenamefont {Dvoretskii},\ and\ \citenamefont
		{Ganichev}}]{Savchenko2018}%
	\BibitemOpen
	\bibfield  {author} {\bibinfo {author} {\bibfnamefont {M.~L.}\ \bibnamefont
			{Savchenko}}, \bibinfo {author} {\bibfnamefont {Z.~D.}\ \bibnamefont {Kvon}},
		\bibinfo {author} {\bibfnamefont {S.}~\bibnamefont {Candussio}}, \bibinfo
		{author} {\bibfnamefont {N.~N.}\ \bibnamefont {Mikhailov}}, \bibinfo {author}
		{\bibfnamefont {S.~A.}\ \bibnamefont {Dvoretskii}}, \ and\ \bibinfo {author}
		{\bibfnamefont {S.~D.}\ \bibnamefont {Ganichev}},\ }\bibfield  {title}
	{\enquote {\bibinfo {title} {Terahertz cyclotron photoconductivity in a
				highly unbalanced two-dimensional electron{\textendash}hole system},}\ }\href
	{\doibase 10.1134/s0021364018160075} {\bibfield  {journal} {\bibinfo
			{journal} {{JETP} Letters}\ }\textbf {\bibinfo {volume} {108}},\ \bibinfo
		{pages} {247--252} (\bibinfo {year} {2018})}\BibitemShut {NoStop}%
\end{thebibliography}

%


\end{document}